\documentclass[aps,prd,twocolumn,showpacs,superscriptaddress,amsmath,graphicx,psfrag,color,longtable]{revtex4}

\usepackage{psfrag}
\usepackage{color}
\usepackage{longtable}

\def\slashed#1{\displaystyle{\not}#1}

\begin{document}

\title{Chiral loop corrections to strong decays of positive and negative 
parity charmed mesons}

\author{Svjetlana Fajfer}
\email[Electronic address:]{svjetlana.fajfer@ijs.si}
\affiliation{J. Stefan Institute, Jamova 39, P. O. Box 3000, 1001 Ljubljana, Slovenia}
\affiliation{Department of Physics, University of Ljubljana, Jadranska 19, 1000 Ljubljana, Slovenia}

\author{Jernej Kamenik}
\email[Electronic address:]{jernej.kamenik@ijs.si}
\affiliation{J. Stefan Institute, Jamova 39, P. O. Box 3000, 1001 Ljubljana, Slovenia}

\date{\today}

\begin{abstract}
We consider chiral loop corrections to the effective couplings which describe 
the strong interactions of two heavy mesons of even or odd parities 
 and a light pseudoscalar meson,  
within a framework which combines 
heavy quark effective theory and chiral perturbation theory. 
The strong couplings 
are extracted  
from the experimental results for the decay widths,  after including
chiral corrections.  
We find that the introduction of positive parity heavy meson fields into the framework gives relevant contributions to the chiral loop corrections of the effective couplings.
The counterterms' contributions are studied by making a randomized fit
distribution of the relevant couplings at $\mu \simeq 1~\mathrm{ GeV}$. Then we discuss the chiral extrapolation based on calculated loop contributions. 
The inclusion of positive parity mesons in the chiral loops is found to be important 
for the lattice QCD studies of charm meson properties, 
where the knowledge of chiral limit behavior is required.
\end{abstract}


\maketitle

\section{Introduction}

In recent years, discoveries of open charm hadrons have stimulated many studies. The first evidence of broad states in the charmed spectrum was provided by the CLEO collaboration~\cite{Anderson:1999wn} which observed a broad $D_1'(2460)$ state with the features of an axial meson. In 2003, Belle~\cite{Abe:2003zm} and Focus~\cite{Link:2003bd} reported the observation of further broad resonances $D_0^{*+}$ and $D_0^{*0}$, ca. $400-500\mathrm{~MeV}$ higher above the usual $D$ states and with opposite parity. In the same year BaBar~\cite{Aubert:2003fg} announced a narrow meson $D_{sJ}(2317)^+$. This was confirmed by 
Focus~\cite{Vaandering:2004ix} and CLEO~\cite{Besson:2003jp} which also 
noticed another narrow state, $D_{sJ}(2463)^+$. Both states were also confirmed by 
Belle~\cite{Krokovny:2003zq}. The measured properties of the $D_{0}^*$ and $D_1'$ states support their interpretation as belonging to the $(0^+,1^+)$ multiplet of $c \bar u$ and $c \bar d$ mesons. Conversely, the $D_{sJ}$ states have been proposed as members of the $(0^+,1^+)$ spin-parity doublet of $c \bar s$ 
mesons~\cite{Bardeen:2003kt,Nowak:2004uv}.
\par
Studies of the basic properties of these states have been triggered particularly 
by the fact that  
the  $D_{sJ}(2317)^+$ and  $D_{sJ}(2463)^+$ states' masses are below threshold for the decay 
into ground state charmed mesons and kaons, as suggested by quark model studies 
\cite{Godfrey:1985xj, Godfrey:1986wj}
and lattice calculations \cite{Hein:2000qu, Dougall:2003hv}.
The strong and electromagnetic transitions of these new states have 
already been studied within a variety of approaches ~\cite{Colangelo:1995ph,Colangelo:1997rp,Colangelo:2003vg,Colangelo:2005gb,Bardeen:2003kt,Mehen:2004uj,Nielsen:2005zr, Lu:2006ry, Wang:2006bs,Wang:2006fg, Wang:2006id}.  
In these investigations heavy hadron chiral perturbation theory (HH$\chi$PT) at leading 
order 
was used as well in attempts to 
explain the observed strong and electromagnetic decay rates~\cite{Colangelo:2003vg,Colangelo:2005gb,Mehen:2004uj}. In ref.~\cite{Mehen:2005hc} 
the authors have considered the leading and subleading ($1/m_H$) operators which describe the even-parity charmed meson finite mass corrections 
within heavy meson chiral perturbation theory. This calculation contained 11 unknown parameters in its predictions. 
After a number of attempts to determine these parameters, the authors of~\cite{Mehen:2005hc} found that a fit to the even-parity masses is possible, provided the couplings describing the strong interaction between even 
and odd parity 
heavy meson states and light pseudoscalars are constrained to lie between 0 and 1. 
\par
In ref.~\cite{Stewart:1998ke} the chiral loop corrections to the $D^* \to D \pi$ and $D* \to D \gamma$ decays were calculated 
and a numerical extraction of the one-loop bare couplings was first performed. Since this calculation preceded the discovery of even-parity meson states,
it did not involve loop contributions containing the even-parity meson states. 
The ratios of the radiative and strong decay widths, and the 
isospin violating decay $D_s^* \to D_s \pi^0$ were used to extract the relevant couplings. 
However, since that time, the experimental situation has improved and therefore we consider   
in this paper the chiral loop contributions to the  
strong decays of the 
even and odd parity charmed meson states using HH$\chi$PT. 
In our calculation we consider the strong 
decay modes given in TABLE~\ref{table_input}.
\begin{table*}
\begin{ruledtabular}
\begin{tabular}{ccccc}
Meson & $J^P$ & Mass~$[\mathrm{GeV}]$ & Width~$[\mathrm{GeV}]$ & $Br.~[\%]$ (final states) \\
\hline
$D^{*+}$~\cite{Eidelman:2004wy} & $1^-$ & $2.010$ & $(9.6 \pm 2.2)\times 10^{-5}$ & $67.7\pm0.5~(D^0 \pi^+)$, $30.7\pm0.5~(D^+ \pi^0)$ \\
$D^{*0}$~\cite{Eidelman:2004wy} & $1^-$ & $2.007$  & $<0.0021$ & $61.9\pm2.9~(D^0 \pi^0)$ \\
$D^{*+}_0$~\cite{Link:2003bd} & $0^+$ & $2.403\pm 0.014 \pm 0.035$ & $0.283 \pm 0.024 \pm 0.034$ & $(D^0 \pi^+)$\footnote{Observed channel.} \\
$D^{*0}_0$ & $0^+$ & $2.350\pm 0.027$\footnote{Average of Belle~\cite{Abe:2003zm} and Focus~\cite{Link:2003bd} values from~\cite{Colangelo:2004vu}.} & $0.262 \pm 0.051^{b}$ & $(D^+ \pi^-)^a$\\
$D^{'0}_1$ & $1^+$ & $2.438\pm 0.030$\footnote{Average of Belle~\cite{Abe:2003zm} and CLEO~\cite{Anderson:1999wn} values from~\cite{Colangelo:2004vu}.} & $0.329 \pm 0.084^{c}$ & $(D^{*+} \pi^-)^{a}$\\
\end{tabular} 
\end{ruledtabular}
\caption{Experimentally measured properties of excited charmed mesons used in our calculations.\label{table_input}}
\end{table*} 
The existing data on the decay widths 
enable us to constrain the leading order parameters: the $D^* D\pi$ coupling $g$, 
$D^{*}_0 D \pi$ coupling $h$, and the coupling $\tilde g$ which enters in the interaction of even parity 
charmed mesons and the light pseudo-Goldstone bosons. Although the coupling 
$\tilde g$ is not yet  experimentally constrained, it moderatelly affects    
the decay amplitudes which we consider in this paper. 
\par
Due to the divergences coming from the chiral loops one needs to include the appropriate counterterms. Therefore we construct a full operator basis of the relevant counterterms and include it into our effective theory Lagrangian. We first systematically calculate the 
chiral corrections to the wave function renormalization and the relevant vertex 
corrections.  
Then we constrain the three strong couplings, first by 
including the loop effects while neglecting finite contributions from the 
counterterms. The reason why one can expect their contributions not to be significant is related to the appropriate choice of the renormalization scale and the fact that the initial and final meson states in processes we consider do not contain strange quarks. Still we study the impact of the counterterms by taking 
their values to be randomly distributed, and observing the results of the couplings' fit. 
\par
A full calculation of the strong decay couplings should contain, in addition to 
the corrections we determine, also
the relevant $1/m_H$ corrections as discussed in ref.~\cite{Mehen:2005hc}. 
However, the number of unknown couplings is yet too high to be determined from 
the existing data. 
Also, recent lattice QCD studies~\cite{Abada:2003un,McNeile:2004rf} of the strong couplings of heavy mesons have noticed that these corrections seem not to be very significant but  pointed out the importance of controlling chiral loop corrections. 
\par
Due to the increase of simulation time, when approaching the chiral limit, lattice studies  use 
large values of the pion mass. In order to connect their results to the physical limit they employ the chiral extrapolation. The knowledge of chiral loop contributions is essential for the stability of this procedure. 
We compare two scenarios of chiral extrapolation within HH$\chi$PT, one including the contribution of the even-parity states, and the other without their contributions.
\par
The paper is organized into the following sections:
Section II contains the description of the framework, Section III is 
devoted to calculations. In Section IV  we explain the extraction of the bare couplings, while Section V contains the study of chiral extrapolations. We conclude with discussion in Section VI of the paper.

\section{Framework}

We use the formalism of heavy meson chiral Lagrangians~\cite{Burdman:1992gh,Wise:1992hn}.  The octet of light pseudoscalar mesons can be encoded into $\Sigma = \xi^2 = \mathrm{exp} (2i \pi^i \lambda^i /f)$ where the $\pi^i \lambda^i$ matrix contains the pseudo-Goldstone fields
\begin{equation}
\pi^i\lambda^i = 
   \begin{pmatrix} \frac{1}{\sqrt 6}\eta + \frac{1}{\sqrt 2} \pi^0 & \pi^+ & K^+ \\
   \pi^- & \frac{1}{\sqrt 6}\eta - \frac{1}{\sqrt 2} \pi^0 & K^0 \\ 
   K^- & \bar K^0 & -\sqrt{\frac{2}{3}}\eta \end{pmatrix}
\end{equation}
and $f\approx 120~\mathrm{MeV}$ at one loop~\cite{Gasser:1984gg}. The heavy-light mesons are customarily cataloged using the total angular momentum of the light degrees of freedom in the heavy meson $j_{\ell}^P$ which is a good quantum number in the heavy quark limit due to heavy quark spin symmetry. The negative $(j_{\ell}^P)=1/2^-$ and positive $(j_{\ell}^P)=1/2^+$ parity doublets can then be respectively represented by the fields $H=1/2(1+\slashed v )[P^*_{\mu} \gamma^{\mu} - P \gamma_5]$, where $P^*_{\mu}$ and $P$ annihilate the vector and pseudoscalar mesons, and $S=1/2(1+\slashed v )[P^*_{1\mu}
\gamma^{\mu}\gamma_{5} - P_0]$ for the axial-vector ($P^*_{1\mu}$) and scalar ($P_0$) mesons.
\par
The Lagrangian relevant for our study of chiral corrections to strong transition among heavy and light mesons is then at leading order in chiral and heavy quark expansion
\begin{eqnarray}
	\mathcal L &=& \mathcal L_{\chi} + \mathcal L_{\frac{1}{2}^-} + \mathcal L_{\frac{1}{2}^+} + \mathcal L_{\mathrm{mix}}, \nonumber\\
	\mathcal L_{\chi} &=& \frac{f^2}{8} \partial_{\mu} \Sigma_{ab} \partial^{\mu} \Sigma^{\dagger}_{ba}, \nonumber\\
	\mathcal L_{\frac{1}{2}^-} &=& - \mathrm{Tr}\left[ \bar H_a (i v \cdot \mathcal{D}_{ab} - \delta_{ab} \Delta_H ) H_b\right] \nonumber\\
	&& + g \mathrm{Tr} \left[ \bar H_b H_a \slashed{\mathcal A}_{ab} \gamma_{5} \right], \nonumber\\
	\mathcal L_{\frac{1}{2}^+} &=& \mathrm{Tr} \left[\bar S_a ( i v \cdot \mathcal{D}_{ab} - \delta_{ab} \Delta_S) S_b \right] \nonumber\\
	&& + \tilde g \mathrm{Tr} \left[ \bar S_b S_a \slashed{\mathcal A}_{ab} \gamma_{5}  \right], \nonumber\\
	\mathcal L_{\mathrm{mix}} &=& h \mathrm{Tr} \left[ \bar H_b S_a \slashed{\mathcal A}_{ab} \gamma_{5}  \right] + \mathrm{h.c.}.
	\label{L_1}
	\label{eq_L_0}
\end{eqnarray}
$\mathcal{D}^{\mu}_{ab} = \delta_{ab}\partial^{\mu} -\mathcal{V}^{\mu}_{ab}$ is the covariant heavy meson derivative. The light meson vector and axial currents are defined as $\mathcal V_{\mu } = 1/2 (\xi^{\dagger} \partial_{\mu} \xi + \xi \partial_{\mu} \xi^{\dagger})$ and $\mathcal A_{\mu } = i/2 (\xi^{\dagger} \partial_{\mu} \xi - \xi \partial_{\mu}\xi^{\dagger})$ respectively. A trace is taken over spin matrices and the repeated light quark flavor indices. Accordingly $\mathcal L_{\chi}$ is of the order $\mathcal O (p^2)$ in the chiral power counting while the rest of this leading order Lagrangian is of the order $\mathcal O(p^1)$.  Exceptions are the $\Delta_H$ and $\Delta_S$ residual masses of the $H$ and $S$ fields respectively. In a theory with only $H$ fields, one is free to set $\Delta_H=0$ since all loop divergences are cancelled by $\mathcal O(m_q)$ counterterms at zero order in $1/m_H$ expansion. However, once $S$ fields are added to the theory, another dimensionful quantity $\Delta_{SH} = \Delta_S - \Delta_H$ enters loop calculations and does not vanish in the chiral and heavy quark limit~\cite{Mehen:2005hc}. It is of the order $\mathcal O(p^0)$ in the standard power counting and in order to keep a well behaved perturbative expansion we need to modify this prescription formally and assign a power counting of $\Delta_{SH}\sim p^1$. At the end of the calculation we fix its value close to the phenomenological mass splitting between the even and odd parity heavy meson multiplets $\Delta_{SH} \approx 400~\mathrm{MeV}$ although smaller values have also been proposed when taking into account next to leading order terms in $1/m_H$ expansion~\cite{Mehen:2005hc}. Conversely, when performing an extrapolation of light quark masses to the chiral limit, we need to treat these non-vanishing $\Delta_{SH}$ contributions carefully.
\par
Following Refs.~\cite{Boyd:1994pa,Stewart:1998ke}, we absorb the infinite and scale dependent pieces from one loop amplitudes into the appropriate counterterms at order $\mathcal O (m_q)$
\begin{eqnarray}
	\mathcal L^{\mathrm{ct}} &=& \mathcal L_{\chi}^{\mathrm{ct}} + \mathcal L_{\frac{1}{2}^-}^{\mathrm{ct}} + \mathcal L_{\frac{1}{2}^+}^{\mathrm{ct}} + \mathcal L_{\mathrm{mix}}^{\mathrm{ct}}, \nonumber\\
	\mathcal{L}_{\chi}^{\mathrm{ct}} &=& \lambda_0 \left[(m_q)_{ab} \Sigma_{ba} + (m_q)_{ab} \Sigma^{\dagger}_{ba}\right], \nonumber\\
	\mathcal{L}_{\frac{1}{2}^-}^{\mathrm{ct}} &=& \lambda_1 \mathrm{Tr}\left[ \bar H_b H_a   (m_q^{\xi})_{ba}\right]  + \lambda'_1 \mathrm{Tr}\left[ \bar H_a H_a (m_q^{\xi})_{bb}\right]\nonumber\\
	&&\hspace{-35pt} + \frac{g \kappa_1}{\Lambda^2_{\chi}} \mathrm{Tr}\left[ (\bar H H \slashed{\mathcal A} \gamma_5)_{ab}(m_q^{\xi})_{ba} \right]  + \frac{g \kappa_3}{\Lambda^2_{\chi}} \mathrm{Tr}\left[ (\bar H H \slashed{\mathcal A} \gamma_5)_{aa}(m_q^{\xi})_{bb} \right] \nonumber\\
	&&\hspace{-35pt} + \frac{g \kappa_5}{\Lambda^2_{\chi}} \mathrm{Tr}\left[ \bar H_a H_a \slashed{\mathcal A}_{bc} \gamma_5 (m_q^{\xi})_{cb} \right] + \frac{g \kappa_9}{\Lambda^2_{\chi}} \mathrm{Tr}\left[ \bar H_c H_a (m_q^{\xi})_{ab} \slashed{\mathcal A}_{bc} \gamma_5 \right] \nonumber\\
	&&\hspace{-35pt} + \frac{\delta_2}{\Lambda_{\chi}} \mathrm{Tr}\left[ \bar H_a H_b i v \cdot \mathcal D_{bc} \slashed{\mathcal A}_{ca} \gamma_5 \right] + \frac{\delta_3}{\Lambda_{\chi}} \mathrm{Tr}\left[ \bar H_a H_b i \slashed{ \mathcal D}_{bc} v \cdot \mathcal A_{ca} \gamma_5 \right] \nonumber\\
	&&\hspace{-35pt} + \ldots,\nonumber\\
	\mathcal{L}_{\frac{1}{2}^+}^{\mathrm{ct}} &=& - \tilde \lambda_1 \mathrm{Tr}\left[  S_a \bar S_b (m_q^{\xi})_{ba}\right]  - \tilde \lambda'_1 \mathrm{Tr}\left[ S_a \bar S_a (m_q^{\xi})_{bb} \right] \nonumber\\
	&&\hspace{-35pt} + \frac{\tilde g \tilde \kappa_1}{\Lambda^2_{\chi}} \mathrm{Tr}\left[ (\bar S S \slashed{\mathcal A} \gamma_5)_{ab}(m_q^{\xi})_{ba} \right]  + \frac{\tilde g \tilde \kappa_3}{\Lambda^2_{\chi}} \mathrm{Tr}\left[ (\bar S S \slashed{\mathcal A} \gamma_5)_{aa}(m_q^{\xi})_{bb} \right] \nonumber\\
	&&\hspace{-35pt} + \frac{\tilde g \tilde \kappa_5}{\Lambda^2_{\chi}} \mathrm{Tr}\left[ \bar S_a S_a \slashed{\mathcal A}_{bc} \gamma_5 (m_q^{\xi})_{cb} \right] + \frac{\tilde g \tilde \kappa_9}{\Lambda^2_{\chi}} \mathrm{Tr}\left[ \bar S_c S_a (m_q^{\xi})_{ab} \slashed{\mathcal A}_{bc} \gamma_5 \right] \nonumber\\
	&&\hspace{-35pt} + \frac{\tilde \delta_2}{\Lambda_{\chi}} \mathrm{Tr}\left[ \bar S_a S_b i v \cdot \mathcal D_{bc} \slashed{\mathcal A}_{ca} \gamma_5 \right] + \frac{\tilde \delta_3}{\Lambda_{\chi}} \mathrm{Tr}\left[ \bar S_a S_b i \slashed{ \mathcal D}_{bc} v \cdot \mathcal A_{ca} \gamma_5 \right] \nonumber\\
	&&\hspace{-35pt} + \ldots,\nonumber\\
	\mathcal L_{\mathrm{mix}}^{\mathrm{ct}} &=& \frac{h \kappa'_1}{\Lambda^2_{\chi}} \mathrm{Tr}\left[ (\bar H S \slashed{\mathcal A} \gamma_5)_{ab}(m_q^{\xi})_{ba} \right] \nonumber\\ 
	&&\hspace{-35pt} + \frac{h \kappa'_3}{\Lambda^2_{\chi}} \mathrm{Tr}\left[ (\bar H S \slashed{\mathcal A} \gamma_5)_{aa}(m_q^{\xi})_{bb} \right] \nonumber\\
	&&\hspace{-35pt} + \frac{h \kappa'_5}{\Lambda^2_{\chi}} \mathrm{Tr}\left[ \bar H_a S_a \slashed{\mathcal A}_{bc} \gamma_5 (m_q^{\xi})_{cb} \right] + \frac{h \kappa'_9}{\Lambda^2_{\chi}} \mathrm{Tr}\left[ \bar H_c S_a (m_q^{\xi})_{ab} \slashed{\mathcal A}_{bc} \gamma_5 \right] \nonumber\\
	&&\hspace{-35pt} + \frac{ \delta'_2}{\Lambda_{\chi}} \mathrm{Tr}\left[ \bar H_a S_b i v \cdot \mathcal D_{bc} \slashed{\mathcal A}_{ca} \gamma_5 \right] + \frac{ \delta'_3}{\Lambda_{\chi}} \mathrm{Tr}\left[ \bar H_a S_b i \slashed{ \mathcal D}_{bc} v \cdot \mathcal A_{ca} \gamma_5 \right] \nonumber\\
	&&\hspace{-35pt} + \mathrm{~h.c.~} + \ldots.\nonumber\\
	\label{eq_L_1}
\end{eqnarray}
Here $\mathcal{D}^{\mu}_{ab} \mathcal A^{\nu}_{bc} = \partial^{\mu} \mathcal A^{\nu}_{ac} + [ \mathcal{V}^{\mu},\mathcal A^{\nu}]_{ab}$ is the covariant derivate acting on the pseudo-Goldstone meson fields and $\Lambda_{\chi}=4\pi f$. Ellipses denote terms contributing only to precesses with more then one pseudo-Goldstone boson as well as terms with $(iv\cdot \mathcal D)$ acting on $H$ or $S$, which do not contribute at this order~\cite{Stewart:1998ke}. The matrix $m_q=\mathrm{diag}(m_u,m_d,m_s)$ induces masses of the pseudo-Goldstone mesons $m^2_{ab}= 4 \lambda_0 (m_a + m_b)/f^2$, where $a,b$ are the light quark flavor indices while $m_q^{\xi} = (\xi m_q \xi + \xi^{\dagger} m_q \xi^{\dagger})$. Thus, in our power counting ($m_q \sim p^2$) all the $\lambda$ terms in $\mathcal L^{\mathrm{ct}}$ are of the order $\mathcal O(p^2)$ while the $\delta$ and $\kappa$ terms are of the order $\mathcal O(p^3)$. Parameters $\lambda'_1$ and $\tilde \lambda'_1$ can be absorbed into the definition of heavy meson masses by a phase redefinition of $H$ and $S$, while $\lambda_1$ and $\tilde \lambda_1$ split the masses of $SU(3)$ flavor triplets of $H_a$ and $S_a$, inducing residual mass terms in heavy meson propagators: $\Delta_a = 2\lambda_1 m_a$ and $\tilde \Delta_a = 2\tilde \lambda_1 m_a$ respectively~\cite{Stewart:1998ke}. As with $\Delta_{H(S)}$, only differences between these  $\mathcal O(p^2)$ residual mass terms enter our expressions. We denote them as $\Delta_{ba}=\Delta_b - \Delta_a$, $\Delta_{\tilde b a}=\tilde \Delta_b - \Delta_a$ and $\tilde \Delta_{ba}=\tilde \Delta_b - \tilde \Delta_a$ and fix them from phenomenological mass splittings between heavy mesons of different $SU(3)$ flavor. Assuming exact isospin and heavy quark spin symmetry, the only nonvanishing splittings are then $\Delta_{3a}\approx\Delta_{\tilde 3 a} \approx \Delta_{3\tilde a}\approx\tilde \Delta_{3a}\approx 100~\mathrm{MeV}$, where here $a=1,2$. For the $\kappa_1$ and $\kappa_9$ terms only the combination $\kappa_{19} = \kappa_1 + \kappa_9$ will enter in an isospin conserving manner here~\cite{Stewart:1998ke} (the $\kappa_1 - \kappa_9$ combination contributes to isospin violating $D_s^*\to D_s\pi^0$ decay, which we do not consider in this paper). In the same manner we only consider contributions of $\kappa'_{19} = \kappa'_1 + \kappa'_9$ and $\tilde \kappa_{19} = \tilde \kappa_1 + \tilde \kappa_9$. At any fixed value of $m_q$, the finite parts of $\kappa_3$, $\tilde \kappa_3$ and $\kappa'_3$ can be absorbed into the definitions of $g$, $\tilde g$ and $h$ respectively~\cite{Stewart:1998ke}. However, one needs to keep in mind that these terms introduce a non-trivial mass dependence on the couplings when chiral extrapolation is considered. The $\delta_2$ and $\delta_3$ enter in a fixed linear combination, introducing momentum dependence into the definition of $g$ of the form $g-(\delta_2+\delta_3) v\cdot k / \Lambda_{\chi}$. For decays with comparable outgoing pseudo-Goldstone energy, this contribution cannot be disentangled from that of $g$~\cite{Stewart:1998ke}. On the other hand, these contributions have to be considered when combining processes with different outgoing pseudo-Goldstone momenta. The same holds for contributions of $\tilde \delta_2$ and $\tilde \delta_3$ in respect to $\tilde g$, as well as $\delta'_2$ and $\delta'_3$ in respect to $h$. At order $\mathcal O(m_q)$ we are thus left with explicit analytic contributions from $\kappa_5$, $\kappa_9$, $\tilde \kappa_5$, $\tilde \kappa_{19}$, $\kappa'_5$, $\kappa'_{19}$, $\delta_2+\delta_3$,$\tilde \delta_2 + \tilde \delta'_3$ and $\delta'_2+\delta'_3$.

%
%
%

\section{Calculation of Chiral Loop Corrections}

Here we present the most important details of our chiral loop calculations. We employ dimensional regularization in the renormalization scheme where the subtracted divergences $2/\varepsilon -\gamma +\log 4\pi + 1$ are absorbed into the appropriate counterterms. 

\subsection{Wavefunction renormalization}

We first calculate the wave function renormalization $Z_{2H}$ of the heavy $H=P,~P^*$ and $P_0$, $P^*_1$ fields. This is done by calculating the heavy meson self-energy $\Pi(v\cdot p)$, where $p$ is the residual heavy meson momentum, and using the prescription
\begin{equation}
Z_{2H} =  1 - \frac{1}{2}\frac{\partial \Pi(v\cdot p)}{\partial v\cdot p}\Big |_{\mathrm{on~mass-shell}},
\end{equation}
where the mass-shell condition is different for the $H$ and $S$ fields due to their residual mass terms $\Delta_{H(S)}$ as well as for the different $SU(3)$ flavors due to the $\Delta_a$ terms. In general it evaluates to $v\cdot p - \Delta_{H(S)}-\Delta_a=0$.
\par
\begin{figure}
\includegraphics{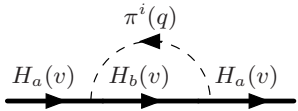}
\caption{\label{diagram_sunrise}Sunrise topology diagram}
\end{figure}

At the $\mathcal O(p^2)$ power counting order we get non-zero contributions to the heavy meson wavefunction renormalization from the self energy ("sunrise") topology diagrams in FIG.~\ref{diagram_sunrise} with leading order couplings in the loop.
\psfrag{pijl}[bc]{$\pi^i(q)$}
\psfrag{Ha}[bc]{$H_a(v)$}
\psfrag{Hb}[bc]{$H_b(v)$}
\psfrag{pi}[bc]{$\pi^i(k)$}
\psfrag{Pa}[bc]{$P_a(v)$}
\psfrag{Pstarb}[bc]{$P^*_b(v)$}
\psfrag{Sb}[bc]{$P_{0b}(v)$}
In the case of the $P$ mesons both vector $P^*$ and scalar $P_0$ mesons can contribute in the loop yielding for the wavefunction renormalization coefficient
\begin{eqnarray}
Z_{2P_a} &=& 1 - \frac{\lambda^i_{ab}\lambda^i_{ba}}{16\pi^2 f^2} \nonumber\\
&& \hskip -1cm \times \left [ 3 g^2 C'_1\left(\frac{\Delta_{ba}}{m_i},m_i\right) - h^2 C'\left(\frac{\Delta_{\tilde b a} + \Delta_{SH}}{m_i} ,m_i\right)\right].\nonumber\\
\label{eq_Z_2_P}
\end{eqnarray}
As in Ref.~\cite{Stewart:1998ke}, a trace is assumed over the inner repeated index(es) (here $b$) throughout the text, while the loop functions $C_i$ and their analytic properties are defined in the Appendix~B. In our power counting scheme, their analytic $\Delta_{ba}$ dependence is of the order $\mathcal O(p^4)$ or higher and could be neglected at this order. However $\Delta_{ba}$ also enter non-analytically and we have to check for sensitivity of our results to these parameters. On the other hand our power counting prescription for $\Delta_{SH}\sim p^1$ leads to well behaved $\mathcal O(p^2)$ and $\mathcal O(p^3)$ analytic contributions to $C_i$.
At leading order in heavy quark expansion, due to heavy quark spin symmetry, the wavefunction renormalization coefficient for the $P^*$ field is identical to that of $P$ although it gets contributions from three different sunrise diagrams with states $P$, $P^*$ and $P_1^*$ in the loops~\cite{Cheng:1993kp}.
\psfrag{Pstara}[bc]{$P^{*\mu}_a(v)$}
\psfrag{Sstarb}[bc]{$P^{*\nu}_{1 b}(v)$}


The positive parity $P_0$ and $P^*_1$ obtain wavefunction renormalization contributions from self energy diagrams (FIG.~\ref{diagram_sunrise}) with $P^*_1$, $P$ and $P_0$, $P^*_1$, $P^*$ mesons in the loops respectively,
\psfrag{Sa}[bc]{$P_{0 a}(v)$}
\psfrag{Sstara}[bc]{$P^{*\mu}_{1 a}(v)$}
which yield identically
\begin{eqnarray}
Z_{2P_{0a}} &=& 1 - \frac{\lambda^i_{ab}\lambda^i_{ba}}{16\pi^2 f^2} \nonumber\\
&& \hskip -1cm \times \left[ 3 \tilde g^2 C'_1\left(\frac{\tilde \Delta_{ba}}{m_i},m_i\right) - h^2 C'\left(\frac{\Delta_{b\tilde a} - \Delta_{SH}}{m_i},m_i\right) \right].\nonumber\\
\end{eqnarray}


\subsection{Vertex corrections}

Next we calculate loop corrections for the $PP^*\pi$, $P_0 P^*_1 \pi$ and $P_0 P\pi$ vertexes. At zeroth order in $1/m_Q$ expansion these are identical to the $P^*P^*\pi$, $P^*_1 P^*_1 \pi$ and $P^*_1P^*\pi$ couplings respectively due to heavy quark spin symmetry. Again we define vertex renormalization factors for on-shell initial and final heavy and light meson fields. Specifically, for the vertex correction amplitude $\Gamma(v\cdot p_i, v\cdot p_f, k_{\pi}^2)$ with heavy meson residual momentum conservation condition $p_f = p_i + k_{\pi}$ one can write the renormalization coefficient $Z_{1 H_i H_f \pi}$ schematically
\begin{equation}
Z_{1 H_i H_f \pi} = 1 - \frac{\Gamma(v\cdot p_i, v\cdot p_f, k_{\pi}^2)}{\Gamma_{l.o.}(v \cdot p_i,v\cdot p_f,k_{\pi}^2)}\Big|_{\mathrm{on~mass-shell}}.
\end{equation}
Here $\Gamma_{l.o.}$ is the tree level vertex amplitude, $p_{i(f)}$ is the residual momentum of the initial (final) state heavy meson $H_{i(f)}$, while $k_{\pi}$ is the pseudo-Goldstone momentum. 
This implies that in the heavy quark limit, one must evaluate the above expression at $k_{\pi}^2 = m_{\pi}^2$, $v\cdot p_{i(f)} = \Delta_{i(f)}$ and consequently $v\cdot k_{\pi} = \Delta_{fi} = \Delta_f - \Delta_i$, where $\Delta_{i(f)}$ is the residual mass of the initial (final) heavy meson fields. While such prescription violates Lorentz momentum conservation at the order $\mathcal O (|\vec k_{\pi}|/m_H) \approx \mathcal O(\sqrt{\Delta_{fi}^2-m_{\pi}^2}/m_H)$, where $\vec k_{\pi}$ is the pion three-momentum, it ensures that in all expressions one only encounters the physical, reparametrization invariant quantities $\Delta_{SH}$, $\Delta_{ab}$, $\Delta_{\tilde a b}$ and $\tilde \Delta_{ab}$~\footnote{This is different from the prescription in Ref.~\cite{Stewart:1998ke}, where $v\cdot k_{\pi}$ entering loop calculations was evaluated as the physical pion energy. However, for the processes considered there, the discrepancy between the two prescriptions is only of the order of a few percent due to the small hyperfine splitting between the relevant initial and final state mesons.}. Satisfying both conditions would require the introduction of two separate velocity scales as in $B \to D$ meson transitions (see e.g.~\cite{Manohar:2000dt}), which is beyond the scope of this work.
\par
\begin{figure}
\includegraphics{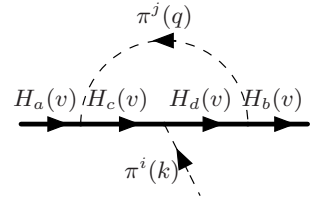}
\caption{\label{diagram_sunrise_road}Sunrise road topology diagram}
\end{figure}

At the $\mathcal O(p^3)$ order, the relevant contributions to the vertexes between the heavy and light mesons come from the "sunrise road" loop topology diagrams in FIG.~\ref{diagram_sunrise_road} with leading order couplings in the loop.
\psfrag{Hc}[bc]{$H_c(v)$}
\psfrag{Hd}[bc]{$H_d(v)$}
\psfrag{pj}[bc]{$\pi^j(q)$}
Contributions from other one loop topology digrams either yield zero due to Lorentz invariance or are cancelled off in our further calculations by the light meson wavefunction renormalization $Z_{2\pi^i}$~\cite{Cheng:1993kp} at the given order. For the case of the $PP^*\pi$ vertex, only $(P^*,P)$, $(P^*,P^*)$ and $(P_0,P^*_1)$ contribute pairwise in the loops. Adding the relevant $\mathcal O(p^3)$ counterterm contributions we thus obtain
\begin{eqnarray}
&&\hspace{-10pt}Z_{1P^*_a P_b\pi^i} = 1 - \frac{	\lambda^j_{ac}\lambda^i_{cd}\lambda^j_{db}}{\lambda^i_{ab} 16\pi^2 f^2} 
\times \Bigg\{ g^2 
			C'_1\left(\frac{\Delta_{ca}}{m_j},\frac{\Delta_{db}}{m_j},m_j\right) 
	\nonumber\\
&&+ \frac{h^2 \tilde g }{g}
		C'\left(\frac{\Delta_{\tilde c a } + \Delta_{SH}}{m_j},\frac{\Delta_{\tilde d b } + \Delta_{SH}}{m_j},m_j\right)  
\Bigg\}
	\nonumber\\
&&+ \frac{\lambda^{i}_{ac}(m_q)_{cb}}{\Lambda_{\chi}\lambda^i_{ab}}(\kappa_{19} + \delta^{ab} \kappa_5) - \frac{\Delta_{ba}}{\Lambda_{\chi}} \frac{\delta_2 + \delta_3}{g}.
\end{eqnarray}
The same expression is obtained for the $P^*P^*\pi$ vertex renormalization from pairs of $(P,P^*)$, $(P^*,P^*)$, $(P^*,P)$ and $(P^*_1,P^*_1)$ running in the loops. 

\par

Similarly for the $P_0 P^*_1 \pi$ and $P^*_1 P^*_1 \pi$ vertexes we get contributions from pairs of $(P_1^*,P_0)$, $(P^*_1,P^*_1)$, $(P,P^*)$ and $(P_0,P^*_1)$, $(P^*_1,P_0)$, $(P^*_1,P^*_1)$, $(P^*,P^*)$ respectively running in the loops
which yield identically
\begin{eqnarray}
&&\hspace{-10pt}Z_{1P^*_{1a} P_{0b}\pi^i} = 1 - \frac{	\lambda^j_{ac}\lambda^i_{cd}\lambda^j_{db}}{\lambda^i_{ab} 16\pi^2 f^2} 
\times \Bigg\{ \tilde g^2   
			 C'_1\left(\frac{\tilde \Delta_{ca}}{m_j},\frac{\tilde \Delta_{db}}{m_j},m_j\right) 
	 \nonumber\\
&& + \frac{h^2 g}{\tilde g} 
		C'\left(\frac{\Delta_{c\tilde a} - \Delta_{SH}}{m_j},\frac{\Delta_{d\tilde b} - \Delta_{SH}}{m_j},m_j\right) 
\Bigg\}
	\nonumber\\
&&+ \frac{\lambda^{i}_{ac}(m_q)_{cb}}{\Lambda_{\chi}\lambda^i_{ab}}(\tilde \kappa_{19} + \delta^{ab} \tilde\kappa_5)- \frac{\tilde \Delta_{ba}}{\Lambda_{\chi}} \frac{\tilde \delta_2 + \tilde \delta_3}{\tilde g}.
\end{eqnarray}
\par

Finally for the $PP_0\pi$ and $P^*P^*_1\pi$ vertexes the pairs of $(P_0,P)$, $(P^*,P^*_1)$ and $(P^*_1,P^*)$, $(P^*,P^*_1)$, $(P,P_0)$ contribute in the loops respectively,
yielding for the renormalization identically
\begin{eqnarray}
&& \hspace{-10pt}Z_{1P_{0a} P_{b}\pi^i} = 1 - \frac{	\lambda^j_{ac}\lambda^i_{cd}\lambda^j_{db}}{\lambda^i_{ab} 16\pi^2 f^2} 
\times \Bigg\{ 
3 g \tilde g 
		C'_1\left(\frac{\tilde \Delta_{ca}}{m_j},\frac{ \Delta_{db}}{m_j},m_j\right) 
\nonumber\\
&& 
- h^2  
			C'\left(\frac{\Delta_{c\tilde a} - \Delta_{SH}}{m_j},\frac{\Delta_{\tilde d b} + \Delta_{SH}}{m_j},m_j\right)
\Bigg\}
	\nonumber\\
&&+ \frac{\lambda^{i}_{ac}(m_q)_{cb}}{\Lambda_{\chi}\lambda^i_{ab}}(\kappa'_{19} + \delta^{ab} \kappa'_5) - \frac{\Delta_{b\tilde a} - \Delta_{SH}}{\Lambda_{\chi}} \frac{\delta'_2 + \delta'_3}{h}.\nonumber\\
\end{eqnarray}


\section{Extraction of bare couplings}

Using known experimental values for the decay widths of $D^{+*}$, $D^{+*}_0$, $D^{0*}_0$ and $D^{'}_1$, and the upper bound on the width of $D^{0*}$ one can extract the values for the bare couplings $g$, $h$ and $\tilde g$ from a fit to the data. The decay rates are namely given by
\begin{equation}
\Gamma(P_a^{*}\to \pi^i P_b) = \frac{|g^{\mathrm{eff.}}_{P^{*}_a P_b\pi^i}|^2}{6 \pi f^2} |\vec k_{\pi^i}|^3,
\end{equation}
and a similar expression (up to polarization averaging phase space factors) for $\Gamma(P_0\to \pi P)$ and $\Gamma(P^*_1\to \pi P^*)$ with $g$ coupling replaced by $h$ and $|\vec k_{\pi}|^3$ replaced by $E^2_{\pi}|\vec k_{\pi}|$. Here $\vec k_{\pi}$ is the three-momentum vector of the outgoing pion and $E_{\pi}$ its energy. The renormalization condition for the couplings can be written as
\begin{equation}
g^{\mathrm{eff.}}_{P_a^*P_b\pi^i} = g \frac{\sqrt{Z_{2P_a}}\sqrt{Z_{2P_b^*}}\sqrt{Z_{2\pi^i}}}{\sqrt{Z_{1P_aP_b^*\pi^i}}} = g Z^g_{P_a^*P_b\pi^i}
\label{eq_g_renorm}
\end{equation}
with similar expressions for the $h$ and $\tilde g$ couplings. 
\par
Due to the large number of unknown counterterms entering our expressions ($\kappa_5$, $\kappa_{19}$, $\tilde \kappa_5$, $\tilde \kappa_{19}$, $\kappa'_5$,  $\kappa'_{19}$, $\delta_2+\delta_3$, $\tilde\delta_2+\tilde\delta_3$ and $\delta'_2+\delta'_3$) we cannot fix all of their values. Therefore we first perform a fit with a renormalization scale set to $\mu\simeq1~\mathrm{GeV}$~\cite{Stewart:1998ke} and we choose to neglect counterterm contributions altogether. Our choice of the renormalization scale in dimensional regularization is arbitrary and depends on the renormalization scheme. Therefore any quantitative estimate made with such a procedure cannot be considered meaningful without also thoroughly investigating counterterm, quark mass and scale dependencies.  
\par
We constrain the range of the fitted bare couplings by using existing knowledge of their dressed values and assuming the first order loop corrections to be moderate and thus also maintaining convergence of the perturbation series:
\begin{itemize}
\item $g$ - following quark model predictions for the positive sign of this coupling~\cite{Becirevic:1999fr} as well as previous determinations~\cite{Colangelo:1995ph, Colangelo:1997rp, Stewart:1998ke, Abada:2003un, McNeile:2004rf,Wang:2006id} we constrain its bare value to the range $g \in [0,1]$.
\item $h$ - this coupling only enters squared in our expressions for the decay rates and was recently found to be quite large~\cite{Colangelo:1995ph, Colangelo:1997rp, Wang:2006bs,Wang:2006id}. We constrain its bare value to the region $|h| \in [0,1]$.
\item $\tilde g$ - non-relativistic quark models predict this coupling to be equally signed and smaller than $g$~\cite{Falk:1992cx}. Similar results were also obtained using light-cone sum rules~\cite{Colangelo:1997rp}, while the chiral partners HH$\chi$PT model predicts $|\tilde g| = |g|$~\cite{Bardeen:2003kt,Nowak:2004uv,Mehen:2005hc}. A recent lattice QCD study~\cite{Becirevic:2005zu} found this coupling to be smaller and of opposite sign than $g$. We combine these different predictions and constrain the bare $\tilde g$ to the region $\tilde g \in [-1,1]$.
\end{itemize} 
We perform a Monte-Carlo randomized least-squares fit for all the three couplings in the prescribed regions using the experimental values for the decay rates from TABLE~\ref{table_input} to compute $\chi^2$ and using values from PDG~\cite{Eidelman:2004wy} for the masses of final state heavy and light mesons. In the case of excited $D^*_0$ and $D'_1$ mesons, we also assume saturation of the measured decay widths with the strong decay channels to ground state charmed mesons and pions ($D^*_0 \to D \pi$ and $D'_1\to D^* \pi$)~\cite{Mehen:2004uj}. 
We obtain the best-fitted values for the couplings $g=0.66$, $|h|=0.47$ and $\tilde g=-0.06$ at $\chi^2=3.9$. The major contribution to the value of $\chi^2$ comes from the discrepancy between decay rates of $D'_1$ and $D_0^*$ mesons. While the former favor a smaller value for $|h|$, the later, due to different kinematics of the decay, prefer a larger $|h|$ with small changes also for the bare $g$ and $\tilde g$ couplings. Similarly, as noted in Ref.~\cite{Becirevic:2004uv}, such differences are due to the uncertainties in the measured masses of the broad excited meson resonances. We expect these uncertainties to dominate our error estimates and have checked that they can shift our fitted values of the bare couplings by as much as $20\%$ depending on which experimental mass values are considered. On the other hand our results are fairly insensitive to the value of the residual mass splittings between heavy fields as $30\%$ variations in $\Delta_{SH}$, $\Delta_{ab}$, $\Delta_{\tilde a b}$ and $\tilde \Delta_{ab}$ only shift our fitted values for the bare couplings a few percent.
We have probed our results to the sensitivity to the renormalization scale $\mu$ and obtained a moderate dependence, namely a $20\%$ variation of scale around $1~\mathrm{GeV}$ results in roughly $10\%$ variation in $g$, $6\%$ variation in $h$ whereas the value of $\tilde g$ is more volatile and can even change sign for large values of $\mu$.
\par 
If we do not include positive parity states' contributions in the loops (and naturally fix $\Delta_H=0$), we obtain a best fit for this coupling $g=0.53$. We see that chiral loop corrections including positive parity heavy meson fields tend to increase the bare $g$ value compared to its phenomenological (tree level) value of $g_{l.o.}=0.61$~\cite{Anastassov:2001cw}, while in a theory without these fields, the bare value would decrease. 
\par
The fitted value of $|h|$, is close to its tree level phenomenological value obtained from the decay widths of $D_0^{*}$ and $D'_1$ mesons (and using the tree level value for $f=130~\mathrm{MeV}$) $h_{l.o}=0.52$.
\par
Our determined magnitude for $\tilde g$ is close to the QCD sum rules determination of its dressed value~\cite{Colangelo:1997rp,Dai:1997df}, but somewhat smaller compared to parity doubling model predictions~\cite{Bardeen:2003kt,Nowak:2004uv}. Its sign is also consistent with the lattice QCD result of Ref.~\cite{Becirevic:2005zu}. Based on this calculation we can derive a prediction for the phenomenological coupling between the heavy axial and scalar mesons and light pseudo-Goldstone bosons $G_{P_{1}^* P_0 \pi}$, which is related to the bare $\tilde g$ coupling as (see e.g. Ref.~\cite{Casalbuoni:1996pg})
\begin{equation}
G_{P_{1}^* P_0 \pi} = \frac{2 \sqrt{m_{P_1^*}m_{P_0}}}{f} \tilde g^{\mathrm{eff.}}_{P_1^* P_0 \pi}.
\end{equation}
Using our best fitted value for $\tilde g = -0.06$ and excited meson masses from TABLE~\ref{table_input}, we predict the absolute value of this phenomenological coupling for the case of $P_1^*=D_1^{'0}$, $P_0=D_0^{*+}$ and $\pi =\pi^-$: $|G_{D_1^{'0} D_0^{*+} \pi^-}| =  6.0$ corresponding to an effective tree level coupling value of $|\tilde g^{\mathrm{eff.}}_{l.o.}|=0.15$.

\par
\begin{figure}
\psfrag{xk}[bc]{$\overline{|\kappa|}$}
\psfrag{yg}[tc][tc][1][90]{$g$}
\psfrag{yh}[tc][tc][1][90]{$|h|$}
\psfrag{ygt}[tc][tc][1][90]{$\tilde g$}
\hspace*{-1cm}\scalebox{0.7}{\includegraphics{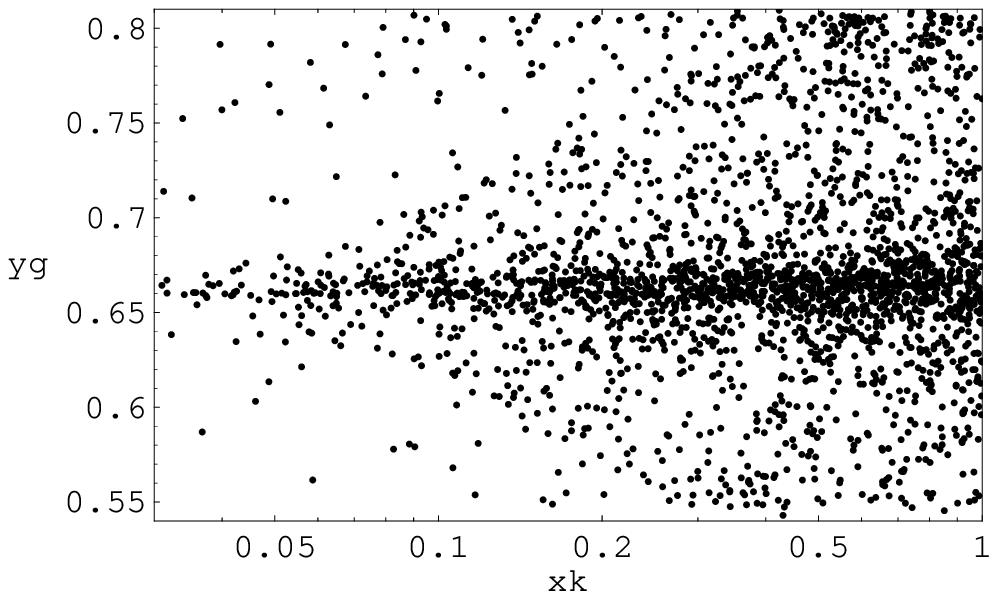}}
\hspace*{-1cm}\scalebox{0.7}{\includegraphics{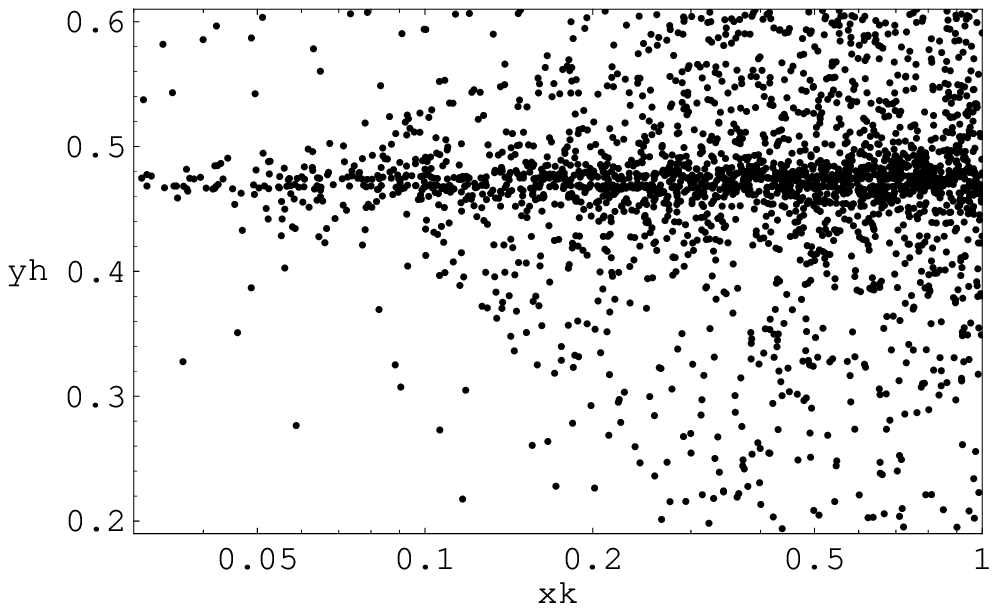}}
\hspace*{-1cm}\scalebox{0.7}{\includegraphics{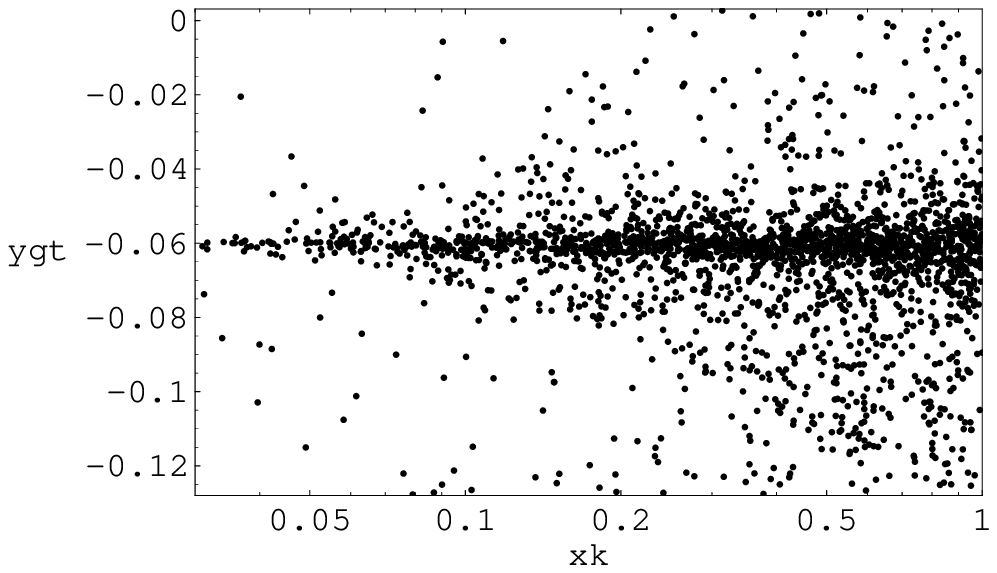}}
\caption{\label{figure_MC}Effect of the $m_q$ and $E_{\pi}$ counterterms of the size order $\overline{|\kappa|}$ on the solutions for the couplings $g$ (top), $|h|$ (middle) and $\tilde g$ (bottom) as explained in the text.}
\end{figure}

Next we study the effects of the counterterms on our couplings fit. Following the approach of Ref.~\cite{Stewart:1998ke} we take the values of $\kappa_5$, $\kappa'_5$, $\kappa_{19}$, $\kappa'_{19}$, $\delta_2+\delta_3$ and $\delta'_2+\delta'_3$ entering our decay modes to be randomly distributed at $\mu\simeq 1~\mathrm{GeV}$ in the interval $[-1,1]$.
Near our original fitted solution, we generate 5000 values of $g$, $|h|$ and $\tilde g$ by minimizing $\chi^2$ at each counterterm sample. For each solution also the average absolute value of the randomized counterterms ($\overline{|\kappa|}$) is computed. We plot the individual coupling solution distributions against this counterterm size measure in FIG.~\ref{figure_MC}. 
We see that the inclusion of counterterms spreads the fitted values of the three couplings. From this we can estimate roughly the uncertainty of the solutions due to the counterterms to be at the one sigma level $g  =  0.66^{ +0.08}_{-0.06}$, $|h|  =  0.47^{ +0.07}_{-0.04}$ and $\tilde g  =  -0.06^{+0.03}_{-0.04}$ if we assume the counterterms do not exceed values of the order $\mathcal O (1)$. This result is in a way complementary to the study of renormalization scale dependence of our couplings' fit. Both are important since although it is always possible in principle to trade the counterterms contributions for a specific choice of the renormalization scale, the latter will be different for different amplitudes where the combination of counterterms will be different.

\section{Chiral Extrapolation}

Next we study the contributions of the additional resonances in the chiral loops to the chiral extrapolations employed by lattice QCD studies to run the light meson masses from the large values used in the simulations to the chiral limit~\cite{Abada:2003un, McNeile:2004rf}.
\par
As already noted in the previous section, the inclusion of heavy excited mesons in the chiral loops introduces large scale dependence into the renormalization of the coupling constants, due to the large splitting between the ground state and excited heavy mesons in the loops. This splitting causes the pseudo-Goldstone bosons in the loops to carry large momenta. They can be highly virtual or, in the cases of $P_0 P \pi$, $P_1^* P^* \pi$ and $P_1^* P_0 \pi$ couplings' renormalizations even real. Such behavior casts doubts on the validity of this extended perturbation scheme, as contributions from higher lying excited heavy meson states seem to dominate the loop amplitudes. The problem can be explored by analyzing the dimensionally regularized loop integral $I_1^{\mu\nu}(m,\Delta)$, which can be found in the Appendix~B. All the other relevant loop integrals can be obtained from this one via algebraic manipulation. The integral is characterized by two dimensionful scales ($m$ and $\Delta$). In addition chiral perturbation theory requires pion momenta (also those integrated over in the loops) to be much smaller than the chiral symmetry breaking scale $\Lambda_{\chi}$. The first integral scale $m$ is the mass of the pseudo-Goldstone bosons running in the loop. In lattice studies, its value is varied and can be taken as large as $m\sim 1~\mathrm{GeV}$. Within chiral perturbation theory however, it is protected by chiral symmetry to be small. On the other hand, once $\Delta$ contains the splitting between heavy meson states of different parity, it is not protected by either heavy quark or chiral symmetries and can be arbitrarily large. Once we attempt to integrate over loop momenta probing also this scale, we are effectively including harder and harder momentum scales in the dimensionally regularized expression as this splitting grows. Finally, as these approach $\Lambda_{\chi}$, the perturbativity and predictability of such scheme break down.
\par
But while the phenomenological couplings' fit seems mainly unaffected by such problems (e.g. the results depend only mildly on the actual value of the mass splitting in the range probed), they play a much more profound role in the chiral extrapolation. The issue is especially severe in the case of pions, which due to their small masses can even become real inside the loops. This introduces uncontrollable final state interactions and causes the chiral extrapolation to diverge. To mend this situation, we attempt on a approximative solution: We expand the integrand of  $I_1^{\mu\nu}(m,\Delta)$ over powers of $\Delta$. We may do this, assuming the relevant loop momentum integration region lies away from the ($v\cdot q - \Delta$) pole, which is true for chiral perturbation theory involving soft pseudo-Goldstone bosons and for a large enough $\Delta$. We obtain a sum of integrals of the form
\begin{eqnarray}
&& I_1^{\mu\nu}(m,\Delta)|_{\Delta = \mathrm{large}} = \nonumber\\
&& = \frac{\mu^{4-D}}{(2\pi)^D} \int \mathrm{d}^D q \frac{q^{\mu} q^{\nu}}{(q^2-m^2)} \frac{-1}{\Delta} (1+\frac{q\cdot v}{\Delta} + \ldots),\nonumber\\
\label{eq_int1}
\end{eqnarray}
where the ellipses denote terms of higher order in the $1/\Delta$ expansion. This greatly simplified integral has a characteristic, that all terms with odd powers of loop momenta in the numerator vanish exactly. Thus, the first correction to the leading $\mathcal O (1/\Delta)$ order truncation appears only at $\mathcal O (1/\Delta^3)$. 
\par
The above described procedure is similar to what is done in the "method of regions" (see e.g.~\cite{Beneke:1997zp}) when one separates out the different momentum scales, appearing in problems involving collinear degrees of freedom. However, here we are only interested in the low momentum part of the whole integral and assume the high momentum contributions are properly accounted for in the counterterms. The leading order term in~(\ref{eq_int1}) then yields for the loop functions
\begin{eqnarray}
C_1(x,m) &=& -\frac{m^3}{4 x} \left[ \frac{m^2}{2} - m^2 \log \left(\frac{m^2}{\mu^2}\right) \right] + \mathcal O \left( \frac{m^3}{x^3} \right), \nonumber\\
C_2(x,m) &=& \mathcal O \left( \frac{m^3}{x^3} \right).
\end{eqnarray}
It is important to stress that the relevant ratio for the validity of this approach is $\Delta/E_{\pi}\gg 1$ as we are expanding in powers of loop momentum, not pseudo-Goldstone masses. 
\par
This approach can alternatively be understood as the expansion around the decoupling limit of the positive parity states with the corresponding contributions being just a series of local operators with $\Delta$ dependent prefactors - effective counterterms of a theory with no positive parity mesons. Therefore any large deviations of this approach from the predictions of a theory without positive parity states and with the couplings properly refitted would signal the breaking of such expansion and the fact that the contributions from "dynamical" positive parity states cannot be neglected.
\par
We compare the above described approximative approach with the complete unaltered expressions for the coupling renormalization from the previous section. As customary we expect the non-analytic chiral log terms to dominate the extrapolation, while any analytic dependence on the pseudo-Goldstone masses can be absorbed into the appropriate counterterms. As an example we write down the dominating contributions to the chiral log extrapolation of the $g$ coupling
\begin{eqnarray}
&& \frac{1}{m_j^2} \frac{\mathrm{d} g^{\mathrm{eff.}}_{P_a^* P_b \pi^i}}{\mathrm{d} \log m_j^2} = \frac{g}{(4\pi f)^2} \nonumber\\
&&\hskip 1cm \times \Bigg\{ \frac{\lambda^j_{ac} \lambda^j_{ca} + \lambda^j_{bc} \lambda^j_{cb}}{2} \left[ -3 g^2 - h^2 \left(1-\frac{6 \Delta_{SH}^2}{m_j^2}\right) \right] \nonumber\\
&&\hskip 1cm + \frac{\lambda^j_{ac}\lambda^i_{cd}\lambda^j_{db}}{\lambda^i_{ab}} \left[ g^2 - h^2 \frac{\tilde g}{g} \left(1-\frac{6 \Delta_{SH}^2}{m_j^2}\right) \right] \Bigg\}.\label{eq_g_extra}
\end{eqnarray}
In the above expression we have for the sake of simplicity neglected the $SU(3)$ flavor splittings between the heavy mesons which are always small compared to $\Delta_{SH}$, are of higher order in the power counting and vanish in the chiral limit. On the other hand one can immediately see, that the $\Delta_{SH}$ contributions due to excited heavy mesons in the loops seemingly dominate the chiral limit, where they diverge. If we instead use the loop integral expansion we effectively replace the $1-6\Delta^2_{SH}/m_j^2$ terms in Eq.~(\ref{eq_g_extra}) with $m^2/4\Delta_{SH}^2$ which on the contrary vanish in the chiral limit. Furthermore, these terms then become of the form $m^4\log m^2$, and thus formally only contribute to the next-to-leading chiral log running. Nontheless, we keep them explicitly in our further calculations to evaluate the size of their contributions. 
\par
In the following quantitative analysis we take the fitted values of the couplings from the previous section. We compare (I) the loop integral expansion, (II) the complete leading log extrapolation with positive parity states contributions, (III) also in the degenerate limit $\Delta_{SH}=0$, and (O) a chiral extrapolation without the positive parity states' contributions and with $g=0.53$ as determined by the coupling fit from the previous section in such a scenario. 
We assume exact $SU(2)$ flavor symmetry and parameterize the pseudo-Goldstone masses according to the Gell-Mann formulae as~\cite{Becirevic:2004uv} 
\begin{equation}
\begin{array}{rclrcl}
m^2_{\pi} & = & \frac{8 \lambda_0 m_s}{f^2} r, & m_K^2 & = & \frac{8 \lambda_0 m_s}{f^2} \frac{r + 1}{2}, \\
m^2_{\eta} & = & \frac{8 \lambda_0 m_s}{f^2} \frac{r+2}{3},
\end{array} 
\end{equation}
where $r=m_{u,d}/m_s$ and $8 \lambda_0 m_s / f^2 = 2m_K^2-m_{\pi}^2 = 0.468~\mathrm{GeV}^2$.
Consequently in the chiral extrapolation we only vary the scale parameter $r$ -- the light quark mass with respect to the strange quark mass which is kept fixed to its physical value. We again choose $\mu=1~\mathrm{GeV}$ for the common renormalization scale. We subtract the counterterm contributions which scale linearly with $r$. The slope of these can be different between the two scenarios and can be inferred from lattice QCD. In Ref.~\cite{Becirevic:2005zu} the $g$ and $\tilde g$ couplings were calculated on the lattice at different $r$ values. However that study used large values of pion masses ($r\sim 1$) where the predispositions for our extrapolation expansion in scenario I are not justified. Also in order to use lattice data to extract the couplings from such a chiral extrapolation one would in addition need lattice results for the $h$ coupling, since it enters in the chiral logs of the other two couplings considered. Instead we normalize our results for the $g$ coupling renormalization in all scenarios at $8 r_{\mathrm{ab}} \lambda_0 m_s / f^2= \Delta_{SH}^2 $ (corresponding in our case to $r_{\mathrm{ab}}=0.34$) to a common albeit arbitrary value of $Z^g_{P_a^*P_b\pi^i}(r_{\mathrm{ab}})=1$ and zero slope. We choose this scale since it is a branching point between extrapolation regions where different approximative limits may apply: by region A we denote the low $r$ range, where our $1/\Delta$ expansion (I) is valid for pion loops; conversely we define region B above this point, where the full contribution (II) and later also the degenerate limit (III) should apply. In order to fit our results to lattice data, one would instead need to add the (counter) terms constant and linear in $r$ to the chiral extrapolation formulae, representing contributions from $s$ and $u,d$ quark masses. Their values could then be inferred together with the values for the bare couplings from the combined fit for all the three couplings to the lattice results.
\par
\begin{figure}
\hspace*{-0.4cm}\scalebox{0.8}{\includegraphics{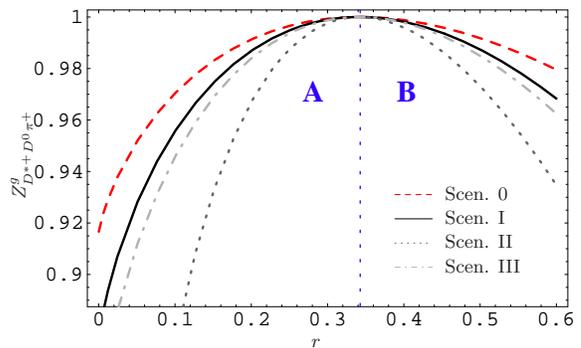}}
\caption{\label{coupling_plot_1}The $g$ coupling renormalization in $D^{*+}\to D^0\pi^+$. Comparison of chiral extrapolation with $g=0.66,~|h| = 0.47,~\tilde g = -0.06$ and (I) loop integral expansion (black, solid), (II) complete log contribution (dark gray, dotted), (III) the degenerate limit (light gray, dash-dotted), and (0) without positive parity doublets included in the loops ($g=0.53,~|h| = 0$) (red, dashed line) as explained in the text.}
\end{figure}

As an example we again consider the strangeless process $D^{*+}\to D^0\pi^+$ 
in FIG.~\ref{coupling_plot_1}. 
\psfrag{xk}[bc]{$r$}
\psfrag{xm}[tc][tc][1][90]{$Z^g_{D^{*+}D^0\pi^+}$}
\psfrag{s0}{Scen. 0}
\psfrag{s1}{Scen. I}
\psfrag{s3}{Scen. II}
\psfrag{s4}{Scen. III}
We can see that including the complete chiral log contributions from excited states in the loops, introduces large ($\gtrsim 30\%$) deviations from the extrapolation without these states. If one instead applies the approximative approach discussed above, the deviations diminish considerably. Surprisingly, the decoupling approach produces results which are numerically very close to the limiting case of degenerate multiplets. One can understand this in terms of the main contributions to the extrapolation still comming from the $g^2$ terms as expected, albeit now with a different fitted value for this coupling. In our approximations the $h^2$ terms then actually reduce this discrepancy due to being oppositely signed. Such corrections to the running due solely to the $h^2$ terms are less then 5\%.   
\par
\begin{figure}
\hspace*{-0.4cm}\scalebox{0.8}{\includegraphics{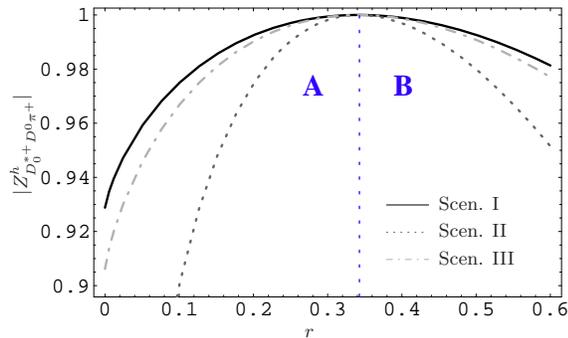}}
\caption{\label{coupling_plot_2} Chiral extrapolation of the $h$ coupling renormalization in $D^{*+}_0\to D^0\pi^+$.
Comparison of chiral extrapolation with $g=0.66,~|h| = 0.47,~\tilde g = -0.06$ and (I) loop integral expansion (black, solid), (II) complete log contribution (dark gray, dotted), and (III) the degenerate limit (light gray, dash-dotted)  as explained in the text. 
}
\end{figure}

For completeness we also plot the chiral extrapolation diagram for the $Z^h_{D^+_0 D^0 \pi^+}$ in FIG.~\ref{coupling_plot_2}.
\psfrag{xk1}[bc]{$r$}
\psfrag{xm1}[tc][tc][1][90]{$|Z^h_{D^{*+}_0 D^0\pi^+}|$}
The results for the chiral extrapolation of the $h$ coupling renormalization are very similar to the $g$ coupling case. Here, in all approximations, the main contributions to the extrapolation come from the $g^2$, $\tilde g^2$ and mixed $g\tilde g$ terms, with smaller corrections due to the $\Delta_{SH}$ dependent $h^2$ terms (except for the complete expression scenario III, where these terms dominate).

\section{Discussion and Conclusion}

Within a HH$\chi$PT  framework, 
which includes even and odd parity heavy meson interactions with light 
pseudoscalars as pseudoGoldstone bosons,  we 
have calculated chiral loop corrections to the strong  $D^* D \pi$ coupling $g$,  
the $D_0^* D \pi$ coupling $h$ and the $D'_1 D_0^* \pi$ coupling
$\tilde g$.
\par
The calculations have been done by considering first the wave functions' renormalization. 
Due to the rather large 
mass splitting between positive in negative parity states $\Delta_{SH}$,
we find that the
perturbation expansion holds for scales below $\mu \leq 1~\mathrm{GeV}$, 
while these new strongly scale dependent corrections become large at higher renormalization scales. 
From the data for the four decay widths (see TABLE~\ref{table_input}) 
we obtain the best-fitted values for the bare couplings, which we summarize in TABLE~\ref{table_summary}. 
\begin{table}
\begin{ruledtabular}
\begin{tabular}{lccc}
 Calculation scheme & $g$ & $|h|$ & $\tilde g$ \\
 \hline
 Leading order & $0.61$~\cite{Anastassov:2001cw} & $0.52$ & $-0.15$\footnote{Effective tree level coupling value derived from one loop calculation for the case $D^{'0}_1 \to D_0^{*+} \pi^-$.}\\
 One-loop without positive parity states & $0.53$ & & \\
 One-loop with positive parity states  & $0.66$ & $0.47$ & $-0.06$ \\
\end{tabular} 
\end{ruledtabular}
\caption{\label{table_summary}Summary of our results for the effective couplings as explained in the text. The listed best-fit values for the one-loop calculated bare couplings were obtained by neglecting counterterms' contributions at the regularization scale $\mu\simeq 1~\mathrm{GeV}$.}
\end{table} 
We are able to determine all the three couplings since the contributions proportional to the coupling $\tilde g$ 
appear  indirectly, through the loop corrections.
One should remember that the quantitatively different results of Ref.~\cite{Stewart:1998ke} appeared before the observation of the even parity heavy meson states and in that study a combination of strong and radiative decay modes was considered in constraining $g$. 
\par
Since we consider decay modes with the  pion in the final state, one should not expect 
sizable  contribution of the counterterms. Namely, the counterterms which appear in our study 
are  
proportional to the light quark masses, and not to the strange quark mass~\cite{Stewart:1998ke}. Nonetheless the effects of counterterm contributions in the decay modes we analyze, 
are estimated by making the random distribution of the relevant 
counterterm couplings. 
The counterterm contributions of order $\mathcal O (1)$ can 
spread the best fitted values of $ g$, $|h|$ by roughly $15\%$ and $\tilde g$ by as much as $60\%$. Similarly, up to $20\%$ shifts in the renormalizations scale modify the fitted values for the $g$ and $|h|$ by less than $10\%$ while $\tilde g$ may even change sign at high renormalization scales. Combined with the estimated $20\%$ uncertainty due to discrepancies in the measured excited heavy meson masses, we consider these are the dominant sources of error in our determination of the couplings. 
One should keep in mind however that without better experimental data and/or lattice QCD inputs, the phenomenology of strong decays of charmed mesons presented above ultimately cannot be considered reliable at this stage.
\par
In the 
work presented in~\cite{Mehen:2005hc}, the next to leading terms 
($1/m_H$) were included in the study of charm meson mass spectrum.  
Due to the very large number of unknown couplings the combination of 
$1/m_H$ and chiral 
corrections does not seem to be possible for the decay modes we consider in 
this paper. However, the studies  of the lattice groups~\cite{Hein:2000qu, Abada:2003un, Dougall:2003hv} indicate 
that the $1/m_H$ corrections do not contribute significantly to 
the their determined values of the couplings, 
and we therefore assume the same to be true in our calculations of chiral corrections. 
\par
Due to computational problems associated with the chiral limit, lattice QCD studies perform calculations at large light quark masses and then employ a chiral extrapolation $m_\pi \to 0$ of their results.
Our analysis of such chiral extrapolation of the coupling $g$ shows that 
the full loop contributions of excited charmed mesons give 
sizable effects in modifying the slope and curvature in the limit $m_\pi \to 0$. We argue that this is due to the inclusion of hard pion momentum scales inside chiral loop integrals containing the large mass splitting between charmed mesons of opposite parity $\Delta_{SH}$ which does not vanish in the chiral limit. If we instead impose physically motivated approximations for these contributions - we expand them in terms of $1/\Delta_{SH}$ - the effects reduce mainly to the changes in the determined values of the bare couplings, used in the extrapolation, with explicit $h$ contributions shrinking to the order of 5\%. Consequently one can infer on the good convergence of the $1/\Delta_{SH}$ expansion.
We conclude that chiral loop corrections in strong charm meson decays can be kept under control, give important contributions and are relevant for 
 the precise extraction of the strong coupling constants $g$, $h$ and $\tilde g$.   

 \begin{acknowledgments}
We thank D. Be\'cirevi\'c for stimulating this study, and for his very relevant comments and suggestions on the first version of the manuscript. We also thank G. Colangelo for fruitful discussions on the topic. Finally we thank the referee of this article for his insightful comments and suggestions. 
This work is supported in part by the Slovenian Research Agency.
\end{acknowledgments}

\appendix

\section{Feynman Rules}

In deriving the Feynman rules we set the overall heavy quark mass scale to a common scale for all processes and states under study inducing a mass gap $\Delta_S$ ($\Delta_H$) terms in the propagators of the positive (negative) parity doublet states due to the relevant residual mass counterterms in the Lagrangian~(\ref{eq_L_0}). The same approach is taken in regards to the $SU(3)$ flavor symmetry breaking contributions, which also induce mass gaps $\Delta_a$ in the heavy meson propagators due to relevant $\mathcal O(m_q)$ counterterm contributions in Lagrangian~(\ref{eq_L_1}). On the other hand, we neglect hyper-fine splittings within individual spin-parity heavy meson doublets. These are degenerate at zeroth order in the $1/m_H$ expansion at which we are working due to heavy quark spin symmetry.
Following is a list of derived Feynman rules used in the calculations in the text. The standard $+i0$-prescriptions are implicitly understood in the propagators.
\begin{longtable}{llcl}
$P_a$ propagator: & \psfrag{P}{$P_a(v)$}\includegraphics{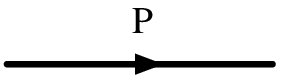} & $=$ & $\frac{i}{2(k\cdot v - \Delta_H -\Delta_a)}$ \\ 
$P^*_a$ propagator: & \psfrag{Pstar}{$P^*_a(v)$}\includegraphics{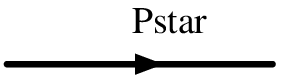} & $=$ & $\frac{-i(g^{\mu\nu}-v^{\mu}v^{\nu})}{2(k\cdot v - \Delta_H -\Delta_a)}$ \\ 
$P_{0a}$ propagator: & \psfrag{S}{$P_{0a}(v)$}\includegraphics{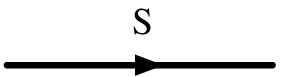} & $=$ & $\frac{i}{2(k\cdot v-\Delta_S-\tilde \Delta_a)}$ \\ 
$P^*_{1a}$ propagator: & \psfrag{Sstar}{$P^*_{1a}(v)$}\includegraphics{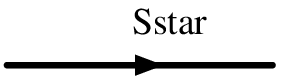} & $=$ & $\frac{-i(g^{\mu\nu}-v^{\mu}v^{\nu})}{2(k\cdot v -\Delta_S - \tilde \Delta_a)}$ \\ 
$\pi^i$ propagator: & \includegraphics{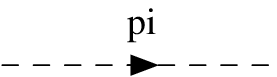} & $=$ & $\frac{i}{k^2-m_i^2}$ \\ 
$P_a P_b^*\pi^i$ coupling: & \includegraphics{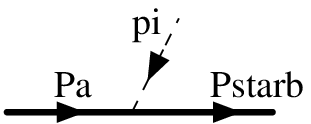} & $=$ & $\frac{2g}{f} k^{\nu} \lambda^i_{ab}$ \\ 
$P^*_a P^*_b \pi^i$ coupling: & \includegraphics{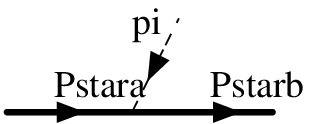} & $=$ & $\frac{2 i g}{f} \epsilon^{\mu\nu\alpha\beta} k^{\alpha} v_{\beta} \lambda^i_{ab}$ \\ 
$P_{0a} P^*_{1b}\pi^i$ coupling: & \includegraphics{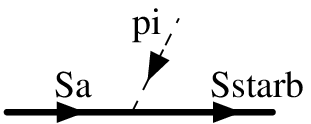} & $=$ & $\frac{2\tilde g}{f} k^{\nu} \lambda^i_{ab}$ \\ 
$P^*_{1a} P^*_{1b} \pi^i$ coupling: & \includegraphics{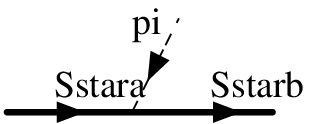} & $=$ & $\frac{2 i \tilde g}{f} \epsilon^{\mu\nu\alpha\beta} k^{\alpha} v_{\beta} \lambda^i_{ab}$ \\ 
$P_a P_{0b}\pi^i$ coupling: & \includegraphics{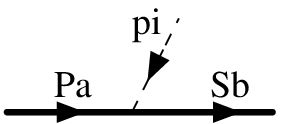} & $=$ & $\frac{-2h}{f} (k\cdot v) \lambda^i_{ab}$ \\
$P^*_a P^*_{1b} \pi^i$ coupling: & \includegraphics{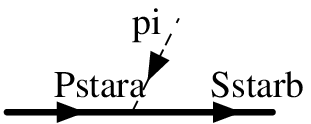} & $=$ & $\frac{2h}{f} (k\cdot v) g^{\mu\nu} \lambda^i_{ab}$ \\
  &  &  \\
\end{longtable} 

\section{Chiral Loop Integrals}

Following is a list of loop integral expressions used in the text. Our notation follows roughly that of Ref.~\cite{Colangelo:1995ph}. All expressions already have the infinite part of the integrals subtracted as explained in the text.
\begin{widetext}
\begin{equation}
I_0(m) = \mu^{(4-D)} \int \frac{\mathrm{d}^D q}{(2\pi)^D} \frac{1}{(q^2-m^2)} = - \frac{i}{16\pi^2} m^2 \log\left(\frac{m^2}{\mu^2}\right),
\end{equation}
\begin{equation}
I_1^{\mu\nu}(m,\Delta) = \mu^{(4-D)} \int \frac{\mathrm{d}^D q}{(2\pi)^D} \frac{q^{\mu} q^{\nu}}{(q^2-m^2)(v\cdot q - \Delta )} = \frac{i}{16\pi^2} \left[ C_1 \left(\frac{\Delta}{m},m\right) g^{\mu\nu} + C_2 \left(\frac{\Delta}{m},m\right) v^{\mu} v^{\nu}\right],
\end{equation}
\begin{equation}
I_2^{\mu}(m,\Delta) = \mu^{(4-D)} \int \frac{\mathrm{d}^D q}{(2\pi)^D} \frac{q^{\mu} }{(q^2-m^2)(v\cdot q - \Delta)} = \frac{i}{16\pi^2}  C \left(\frac{\Delta}{m},m\right) \frac{v^{\mu}}{\Delta},
\end{equation}
\begin{eqnarray}
I_3^{\mu\nu}(m,\Delta_1,\Delta_2) &=& \mu^{(4-D)} \int \frac{\mathrm{d}^D q}{(2\pi)^D} \frac{q^{\mu} q^{\nu}}{(q^2-m^2)(v\cdot q - \Delta_1)(v\cdot q - \Delta_2)}\nonumber\\
&=& \frac{1}{\Delta_1-\Delta_2} \left[ I_1^{\mu\nu}(m,\Delta_1)  - I_1^{\mu\nu}(m,\Delta_2) \right],
\end{eqnarray}
where
\begin{equation}
I_3^{\mu\nu}(m,\Delta,\Delta) = \frac{\mathrm{d}}{\mathrm{d}\Delta} I_1^{\mu\nu}(m,\Delta).
\end{equation}
In the text we then make use of the following expressions
\begin{equation}
C(x,m) =\frac{m^3}{9} \left[ -18x^3 + (18x^3 -9x) \log\left(\frac{m^2}{\mu^2}\right) + 36 x^3 F\left(\frac{1}{x}\right) \right],
\end{equation}
\begin{equation}
C_1(x,m) =\frac{m^3}{9} \left[ -12x + 10 x^3 + (9x-6x^3) \log\left(\frac{m^2}{\mu^2}\right) - 12 x (x-1) F\left(\frac{1}{x}\right)  \right],
\end{equation}
\begin{equation}
C_2(x,m) = C(x,m) - C_1(x,m),
\end{equation}
with
\begin{equation}
C'_{1,2} \left(x,y,m\right) = \frac{1}{m}\frac{1}{x-y} [C_{1,2}(y,m)-C_{1,2}(x,m)],
\end{equation}
\begin{equation}
C'_{1,2}(x,m)  = C'_{1,2} \left(x,x,m\right) = \frac{1}{m}\frac{\mathrm{d}}{\mathrm{d}x} C_{1,2}(x,m).
\end{equation}
The function $F(x)$ was calculated in Ref.~\cite{Stewart:1998ke}
\begin{equation}
F\left(\frac{1}{x}\right) = \left\{ 
\begin{array}{lc}
\frac{\sqrt{x^2-1}}{x} \log\left( x+\sqrt{x^2-1} \right), & |x| \geq 1, \\
-\frac{\sqrt{1-x^2}}{x} \left[\frac{\pi}{2} - \tan^{-1}\left( \frac{x}{\sqrt{1-x^2}} \right) \right], & |x| \leq 1.
\end{array} 
\right.
\end{equation}

\end{widetext}

\bibliography{article}

\end{document}